\documentclass[twocolumn,showpacs,preprintnumbers,amsmath,amssymb]{revtex4}%
\usepackage{graphicx}
\usepackage{dcolumn}
\usepackage{bm}
\usepackage{color}
\usepackage{amsmath}
\usepackage{amsfonts}
\usepackage{amssymb}%
\begin{document}
\title{The effect of Coulomb field on laser-induced ultrafast imaging methods}
\author{XiaoLei Hao$^{1}$}
\author{YuXing Bai$^{1}$}
\author{XiaoYun Zhao$^{1}$}
\author{Chan Li$^{1}$}
\author{JingYu Zhang$^{1}$}
\author{JiLing Wang$^{1}$}
\author{WeiDong Li$^{1}$}
\email{wdli@sxu.edu.cn}
\author{ChuanLiang Wang$^{2,5}$}
\author{Wei Quan$^{2}$}
\author{XiaoJun Liu$^{2}$}
\email{xjliu@wipm.ac.cn}
\author{Zheng Shu$^{3}$}
\author{Mingqing Liu$^{3}$}
\author{Jing Chen$^{3,4}$}
\email{chen_jing@iapcm.ac.cn}
\affiliation{$^{1}$Institute of Theoretical Physics and Department of Physics,State Key
Laboratory of Quantum Optics and Quantum Optics Devices,Collaborative
Innovation Center of Extreme Optics, Shanxi University,Taiyuan 030006, China}
\affiliation{$^{2}$State Key Laboratory of Magnetic Resonance and Atomic and Molecular Physics, Wuhan Institute of Physics and Mathematics, Innovation Academy for Precision Measurement Science and Technology, Chinese Academy of Sciences, Wuhan 430071, China}
\affiliation{$^{3}$Institute of Applied Physics and Computational Mathematics,P.O. Box
8009, Beijing 100088, China}
\affiliation{$^{4}$Center for Advanced Material Diagnostic Technology, Shenzhen Technology
University, Shenzhen 518118, China.}
\affiliation{$^{5}$Present address: School of Physics and Electrical Engineering, Kashgar
University, Kashgar 844006, China}
\date{\today}

\begin{abstract}

By performing a joint theoretical and experimental investigation
on the high-order above-threshold ionization (HATI) spectrum, the
dominant role of the 3rd-return-recollision trajectories in the
region near the cutoff due to the ionic Coulomb field is
identified. This invalidates the key assumption adopted in the
conventional laser-induced electron diffraction (LIED) approach
that the 1st-return-recollision trajectories dominate the spectrum
according to strong field approximation (SFA). Our results show
that the incident (return) electron beams produced by the 1st and
3rd returns possess distinct characteristics of beam energy, beam
diameter and temporal evolution law due to the influence of
Coulomb field, and therefore the extracted results in the LIED
will be altered if the significance of the 3rd-return-recollision
trajectories is properly considered in the analysis. Such Coulomb
field effect should be taken into account in all kinds of
laser-induced imaging schemes based on recollision.

\end{abstract}

\pacs{33.80.Rv}
\maketitle

As one of the most important processes in strong field physics,
recollision has provided an unprecedented insight into the inner
working of atoms and molecules~\cite{CorkumPT2011,JPB}. In
recollision picture, one electron is liberated from the target
atom or molecule through tunneling ionization and then be
accelerated in the field and pulled back by the field to collide
with the parent ion. Most intriguing phenomena in strong field
physics, such as high-order above-threshold ionization(HATI), high
harmonics generation(HHG) and nonsequential
double-ionization(NSDI), can be well understood based on the
recollsion physics (see, e.g.,
~\cite{Proto1997,BackerOP2002,KrauszPHYS2009,BackerPHYS2012} for
reviews and references therein).

\begin{figure}[ptb]
\centering
\includegraphics[width=3.4in]{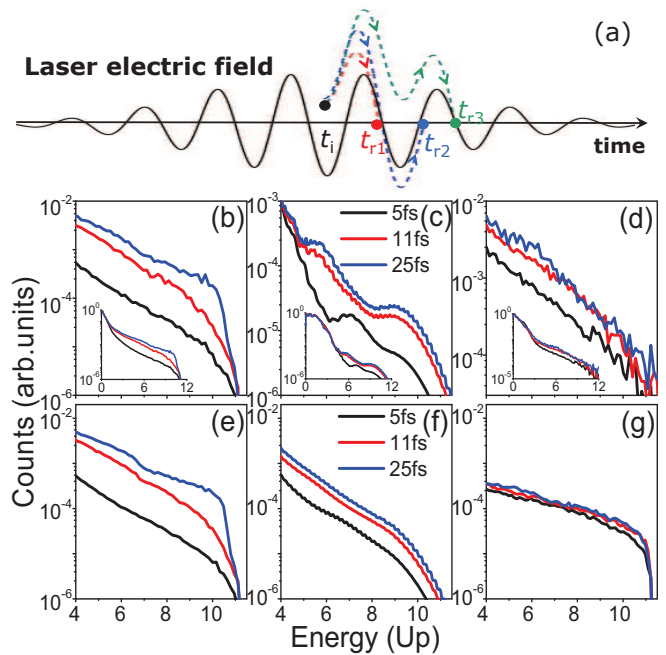}\caption{(Color online) (a) Sketch map
to illustrate the multiple-return-recollision trajectories. For a
multicycle laser field, after the electron is ionized through
tunneling at $t_{i}$, it may come back to collide with the core
upon the first return at $t_{r1}$ or miss the core and collide
with it at $t_{r2}$, $t_{r3}$ ... (b)-(g) Simulated and measured
ATI spectra for Ar atoms exposed to laser pulses with different
durations at peak intensity of $1.25\times10^{14}$W/cm$^{2}$ and
center wavelength of 800 nm. (b)-(d) show CTMC, TDSE and measured
high energy part of the spectra in the direction of polarization,
respectively. The insets present the entire spectra. (e) and (f)
show angle-integrated spectra calculated by CTMC and TDSE,
repectively. (g) shows the angle-integrated spectra calculated
with CTMC by employing the Yukawa potential in the evolution of
electrons, while all other spectra in Fig. \ref{FIG.1} are
calculated in hydrogen-like Coulomb
potential.}%
\label{FIG.1}%
\end{figure}

Since the products upon recollision carry information of the
parent ion, they can be used to probe the structure of the parent
ion. Various methods have been proposed to image molecules in
intense laser fields based on the analysis of different products
upon recollision, such as laser-induced electron diffraction
(LIED)~\cite{MmSCI2008,BlagaNAT2012,XuNC2014,PullenNC2016,MgNC2015,BwSCI2016},
molecular clock\cite{HnNAT2002,HnNAT2003,XTongPRL2003,AsPRL2004},
tomographic imaging of
orbitals~\cite{ItataniNAT2004,HawsslerPHYS2010} and laser-induced
inelastic diffraction (LIID)~\cite{QuanPRL2017}. Since these
methods are self-imaging approaches based on coherent electron
scattering, they can provide an unprecedented spatial-temporal
resolution. In general, as long as the imaging method is based on
electron scattering, the resolution of extracted results will
depend upon the parameters of the incident electron beam, such as
the beam energy, the beam diameter and the temporal evolution of
the beam. Unlike the conventional electron diffraction (CED)
method in which the information of the incident electron beam is
well known, the incident (return) electron beam in the
self-imaging method is produced by the laser-induced ionization of
the target itself and its information is much more complicated. In
LIED, for example, the temporal resolution relies on the knowledge
of the temporal evolution of the beam, and more specifically,
deconvolution of the exact recollision time. In the recollision
process, however, the electron may miss the parent ion at the 1st
return but collide with it at the subsequent returns as
illustrated in Fig. \ref{FIG.1} (a), which results in a large
uncertainty in the recollision time. To complicate matters
further, the incident energy and the impact parameter for
different returns vary. In LIED, the aforementioned complexity is
largely ignored by applying the strong field approximation (SFA)
wherein the Coulomb interaction between the parent ion and the
freed electron is ignored. According to SFA, the maximal kinetic
energy at collision for 1st-return-recollision trajectories is
much higher than that for multiple-return-recollision
trajectories. Thus, the high energy part of the photoelectron
momentum distribution selected for analysis in LIED is assumed to
be produced mostly by the 1st-return-recollision trajectories.
Furthermore, due to the spread of the electron wavepacket, the
probability of the multiple-return-recollision trajectories is
much smaller.

Nevertheless, more and more studies have shown that the Coulomb field plays an
important role in the ionization dynamics of atoms and molecules in intense
laser
fields\cite{BlagaNPH2009,QuanPRL2009,WuPRL2012,QuanREP2016,ShafirPRL2013,LiPRL2013}%
. Specially, the Coulomb focusing effect will significantly
improve the contribution of the multiple-return-recollision
trajectories~\cite{BrabecPRA1996,YudinPRA2001} and in some cases
it will even exceed that of the 1st
return~\cite{XTongPRL2003,HaoPRA2011,JiaPRA2013} in inelastic
recollision process. Therefore, the basic assumption on the
incident electron beam in the ultrafast imaging method may be in
question, and, the impact of the Coulomb field on the imaging
results has to be carefully accessed. In this paper, we
theoretically and experimentally investigate the plateau in ATI
spectrum and a distinct laser pulse duration dependence of the
plateau is revealed. Analysis shows that this dependence is
closely related to the increasing contribution of the
3rd-return-recollision trajectories due to the effect of Coulomb
field. For multi-cycle laser field, this contribution becomes even
dominant in the near-cutoff HATI spectrum. Since electron beams
associated with the 1st and 3rd returns show distinct time
evolution characteristic and impact parameter distribution, using
the 3rd-return-recollision trajectories for analysis in LIED will
change the imaging results.

In Fig. \ref{FIG.1} we present the photoelectron energy spectra
calculated with classical-trajectory Monte Carlo method (CTMC)
(Fig. \ref{FIG.1}(b)) and numerical solution of time-dependent
Schr\"{o}dinger equation (TDSE) (Fig. \ref{FIG.1}(c)), and also
the measured spectra (Fig. \ref{FIG.1}(d)), for Ar atoms exposed
to intense laser pulses with different pulse durations at peak
intensity of $1.25\times10^{14}$W/cm$^{2}$ and center wavelength
of 800 nm. The details of the CTMC method
\cite{QuanREP2016,HuPLA1997,ChenPRA2000} are presented in the
supplementary material. The TDSE is solved using the freely
available software QPROP \cite{QPROP}. In both calculations, the
hydrogen-like Coulomb potential is applied. The linearly polarized
laser field has a
sine-square pulse envelope in the form of $E\left(  t\right)  =E_{0}%
\sin^{2}\left(  1.14 t/\tau_{p}\right)  \cos\left(  \omega
t+\varphi_{0}\right) $, where $\varphi_{0}$ is the
carrier-envelope phase (CEP), $\tau_{p}$ is the duration of the
pulse. Here the pulse duration is defined as the full width at
half maximum (FWHM) of the intensity. The measured results are
obtained with CEP unlocked, and correspondingly, in calculation of
CTMC each trajectory is calculated with random CEP, and in TDSE
the spectra are averaged over different CEPs with regular interval
of $\pi/8$.

In our experiment (see supplementary material for details), the
laser beam is introduced into the vacuum chamber of a homemade
time-of-flight (TOF) photoelectron kinetic energy spectrometer
\cite{QuanPRA2013,LaiJPB2018} with a limited detection angle of
0.026 sr (opening angle of $5{{}^\circ}$) in the direction of
laser polarization. To compare directly with the measured spectra
in Fig. \ref{FIG.1}(d), only trajectories with momentum direction
in the corresponding range of $\theta_{p}<5{{}^\circ}$, where
$\theta_{p}$ is the angle between the final momentum and the laser
polarization, are considered to obtain the CTMC spectra (Fig.
\ref{FIG.1}(b)). For TDSE, in Fig. \ref{FIG.1}(c) we present the
spectra along the direction of the laser
polarization, i.e. $\left(  \mathbf{dw/dEd\Omega}\right)  \left\vert _{\mathbf{\theta=\varphi=0}%
}\right.  $. The ATI peaks in TDSE spectra are smoothed over by
averaging adjacent data points to make the variation of the
plateau more visible. All spectra in Fig. \ref{FIG.1} are
normalized to themselves by dividing the maximum of the individual
spectrum. The simulated and measured entire spectra (insets in
Figs. \ref{FIG.1}(b)-(d)) exhibit the well-documented ATI spectral
features, i.e., a rapid decrease within $2U_{p}$ followed by a
plateau extending to $10U_{p}$. If we focus on the dependence of
the plateau on pulse duration, a qualitative agreement can be
found between simulations (Figs. \ref{FIG.1}(b), (c)) and
measurement (Fig. \ref{FIG.1} (d)). The yield of the plateau first
increases quickly when pulse duration increases from 5 fs to 11
fs, then only increases slightly until the pulse duration
increases to 25 fs \cite{50fs}. In Figs. \ref{FIG.1}(e) and (f),
we also present the angle-integrated spectra calculated by CTMC
and TDSE to show that the pulse duration dependence is also
prominent in the case of high-acceptance-angle with which LIED
measurement is performed. This is not surprising, because the
majority of photoelectrons will move along the polarization
direction in linearly polarized laser field.

In the following we try to understand the pulse duration
dependence of the plateau with CTMC approach by taking advantage
of the transparent intermediate process that can be explored with
this approach. If we employ the Yukawa potential instead of the
Coulomb potential in the evolution of electrons after leaving the
tunnel exit in CTMC, the spectra become independent of the pulse
duration as shown in Fig. \ref{FIG.1}(g). The Yukawa potential is
of the form $V\left( r\right)  =-\left( Z^{\prime}/r\right)
\exp\left(  -r/a\right) $, with parameters $Z^{\prime}=4.547$ and
$a=4$ which are chosen to retain the ground-state energy of Ar by
TDSE. Therefore, the pulse duration dependence of the plateau
comes from the effect of the ionic Coulomb potential.

\begin{figure}[ptb]
\centering
\includegraphics[width=3.3in]{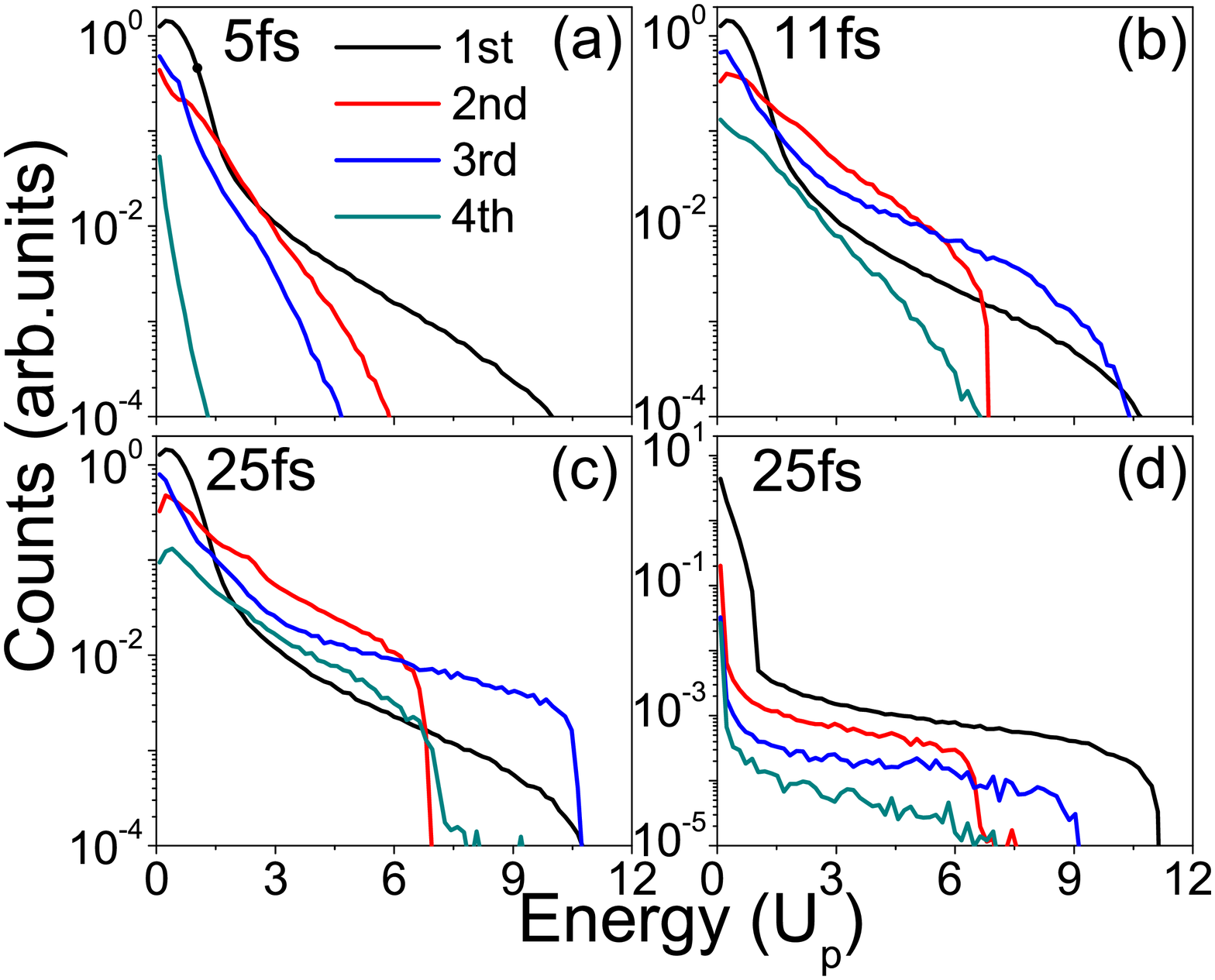}\caption{(Color online) CTMC calculated
spectra with angle integrated for different returns of recollision
trajectories for Ar atoms in laser pulses at peak intensity of
$1.25\times10^{14}$W/cm$^{2}$ and wavelength of 800 nm. (a)-(c)
Spectra calculated using hydrogen-like Coulomb potential for
different pulse durations as indicated in the figures. (d) Spectra
calculated using Yukawa potential for pulse duration of 25 fs.}%
\label{FIG.2}%
\end{figure}

Next we will show how the Coulomb field affects the plateau of the
ATI spectrum. The pulse duration dependence of the plateau in Fig.
\ref{FIG.1} indicates that some specific
multiple-return-recollision trajectories contribute significantly
to the plateau. In the CTMC model, the different return
recollision trajectories can be distinguished according to the
travel time $t_{t}$ defined as the interval between the ionization
time and the recollision time. For trajectories with $t_{t}$ in
the interval [$(n/2)T,~((n+1)/2)T$] ($T$ is the optical cycle), we
denote them as the $n$th-return-recollision trajectories
\cite{HaoPRA2011}. The contributions of different returns to the
spectra in Fig. \ref{FIG.1}(b) are presented in Fig.
\ref{FIG.2}(a)-(c). In the case of the shortest duration of 5 fs
(Fig. \ref{FIG.2}(a)), the 1st-return-recollision trajectories
play a dominant role in the plateau, while the contributions of
the multiple-return-recollision trajectories can be ignored. When
the pulse duration increases to 11 fs (Fig. \ref{FIG.2}(b)), the
contributions of the multiple-return-recollision trajectories
increase significantly. The yields of the 2nd and 3rd returns even
exceed that of the 1st return. Consequently, the
2nd-return-recollision trajectories becomes dominant in the low
energy part of the plateau while the 3rd-return-recollision
trajectories dominate the high energy part. When the pulse
duration further increases to 25 fs (Fig. \ref{FIG.2}(c)), the
contributions of the multiple returns continue to increase but
much more slightly compared with that from 5 fs to 11 fs. The
cutoffs of multiple-return-recollision trajectories, especially
the 3rd and 4th return trajectories, also increase with pulse
duration, which can be attributed to the variation of the pulse
envelope. Since electrons are mostly probably ionized around the
maximum of the envelope, the laser field of 5 fs pulse will
decrease dramatically when electrons come back to collide with the
ions due to steep gradient of the pulse envelope. So the energy
cutoffs are much smaller than that in the plane-wave laser field
\cite{BeckerPHYS1994,HaoCPL2016}. When the pulse duration
increases, the cutoffs increase due to the smaller gradient of
pulse envelope. At 25 fs, the cutoff of the 3rd return even
becomes equal to the 1st return.

However, according to SFA, trajectories with longer travel time
will have smaller probability to collide with the ion due to wave
packet spreading. So the multiple-return-recollision trajectories
should have smaller contributions to the plateau, which is induced
by backscattering, than the 1st-return-recollision trajectories.
This can be clearly seen in the results of Yukawa potential in
Fig. \ref{FIG.2}(d), in which the contribution of the 1st return
is higher than that of multiple returns. In addition, the cutoff
of the 3rd return is significantly lower than the 1st return for
Yukawa potential. Therefore, the effect of Coulomb field increases
not only the contribution of the 3rd return at the high energy
part of the plateau but also the energy cutoff. This is the reason
why the plateau exhibits distinct dependence on the pulse duration
in Fig. \ref{FIG.1}.

\begin{figure}[ptb]
\centering
\includegraphics[width=3.5in]{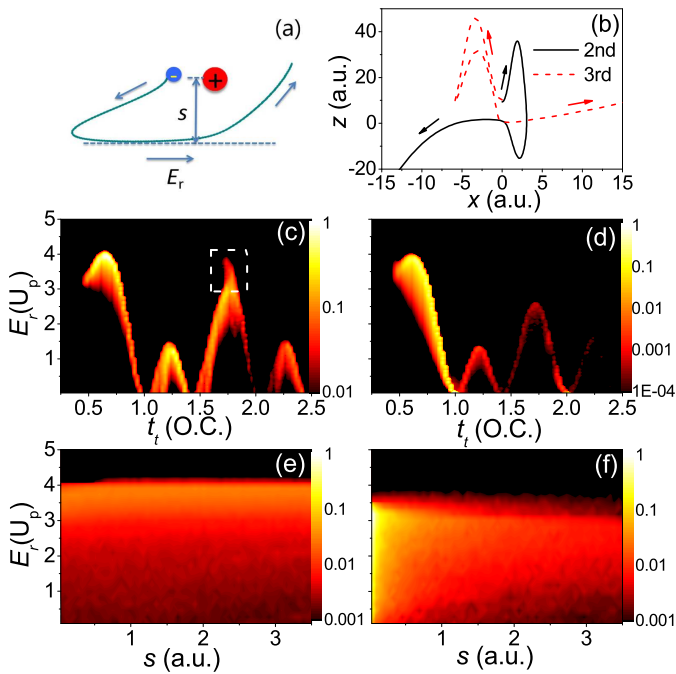}\caption{(Color online) (a)
Illustration of the recollision energy $E_{r}$ and the impact
parameter $s$. (b) Typical 2nd- and 3rd-return-recollision
trajectories contributing to the high energy part of the plateau
in ATI spectrum. See text for details. (c), (d) 2D distributions
of recollision energy $E_{r}$ and travel time $t_{t}$ for Coulomb
potential and Yukawa potential, respectively. The travel time
$t_{t}$ is defined as the interval between the ionization time and
the recollision time. See text for the white dashed line box in
(c). (e), (f) 2D distributions of recollision energy $E_{r}$ and
impact parameter $s$ for the 1st- and 3rd-return-recollision
trajectories, respectively. The distributions are normalized by
dividing the maximum of the 3rd return. The laser parameters are
$1.25\times10^{14}$W/cm$^{2}$, 800 nm, and 25 fs.}%
\label{FIG.3}%
\end{figure}

The dominant contribution of the 3rd return to the spectrum near
the cutoff as well as the increase of the cutoff energy can be
attributed to modification of the incident (return) electron beam
before recollision in the ionic Coulomb potential. The electron
beam can be characterized by the temporal evolution of the beam
(travel time $t_{t}$), the beam energy (recollision energy
$E_{r}$) and the beam diameter (impact parameter $s$) as
schematically illustrated in Fig. \ref{FIG.3}(a). In Fig.
\ref{FIG.3}(c)-(d) we present the CTMC calculated 2D distributions
of $E_{r}$ and travel time $t_{t}$ for Coulomb potential and
Yukawa potential, respectively, at pulse duration of 25 fs. The
four peaks in each plot correspond directly to the four returns,
respectively. The distributions in Figs. \ref{FIG.3}(c) and (d)
are similar except for some details. First, the relative
contributions of multiple returns to the 1st return are much
smaller in the case of Yukawa potential. Second, in Fig.
\ref{FIG.3}(c) there is a peak on the top of the 3rd return
(indicated by dashed line box), while it is absent in the
distribution of Yukawa potential in Fig. \ref{FIG.3}(d). With this
peak the recollision energy cutoff of the 3rd return even extends
to a value very close to the 1st return as shown in Fig.
\ref{FIG.3}(c). This causes the final energy cutoff of the 3rd
return to be identical to the 1st return as shown in Fig.
\ref{FIG.2}(c). After making analysis on the trajectories
contributing to the peak on the top of the 3rd return, we find
that more than $99\%$ of the trajectories always move on one side
of the $z=0$ plane (laser field is polarized along $z$ axis)
before recollision. The typical 3rd return trajectory is shown in
Fig. \ref{FIG.3}(b). For such a trajectory, the Coulomb force on
the electron is always in the same direction as the momentum
before recollision, which results in an increase of the momentum
at recollision. While for the 2nd return and 4th return
trajectories, the electrons leave the tunnel exit on one side of
the ions but will return to the ions from the other side, so they
have to cross the $z=0$ plane to scatter hardly with the ions to
produce high energy electrons. As a result, the Coulomb force will
reverse its direction correspondingly, thus the net effect of
Coulomb field is negligible. The typical 2nd-return-recollision
trajectory is also shown in Fig. \ref{FIG.3}(b). After making
statistics on trajectories contributing to the high energy part of
the plateau ($E>5U_{p}$) in Fig. \ref{FIG.2}(c), we find that
almost $100\%$ of the 2nd- and 4th-return-recollision trajectories
cross the $z=0$ plane before recollision, while the proportion for
the 3rd return is only $50\%$.

\begin{figure}[ptb]
\centering
\includegraphics[width=3.5in]{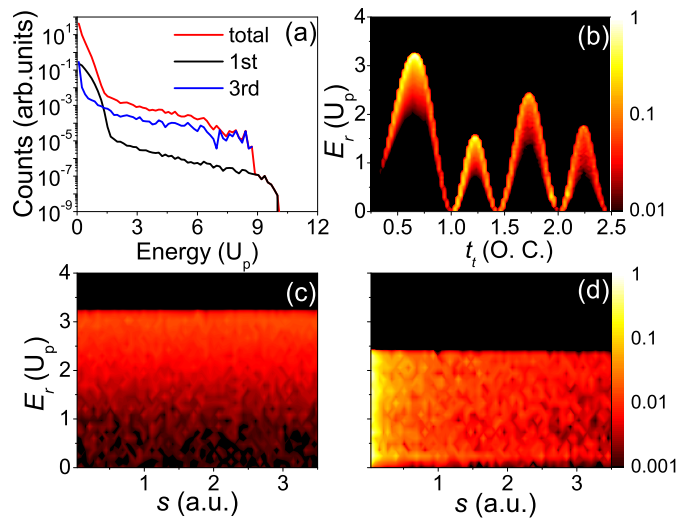}\caption{(Color online) CTMC calculated
results for hydrogen-like Ar atoms in laser pulses at peak
intensity of $1.25\times10^{14}$W/cm$^{2}$, wavelength of $3100$nm
and pulse duration of 100 fs. (a) ATI spectra with angle
integrated. (b) 2D distribution of recollision energy $E_{r}$ and
travel time $t_{t}$. (c), (d) 2D distributions of recollision
energy $E_{r}$ and impact parameter $s$ for the 1st and 3rd
returns, respectively. The
distributions are normalized by dividing the maximum of the 3rd return.}%
\label{FIG.4}%
\end{figure}

In Figs. \ref{FIG.3}(e) and \ref{FIG.3}(f) we plot the 2D
distribution for the recollision energy $E_{r}$ and the impact
parameter $s$ for the 1st and the 3rd return, respectively (see
supplementary material for details of calculating $s$). The
distributions are normalized to the maximum of the 3rd return to
underline the relative contribution between the two returns. The
distribution of the 1st return is more concentrated in the high
energy region near the maximal return energy but distributes
rather uniformly in the impact parameter direction. In contrast,
the distribution of the 3rd return is rather uniform in the energy
axis but concentrates in the small impact parameter regime.
Therefore, the 3rd-return-recollision trajectories are more likely
to experience backscattering to produce high energy electrons, and
hence its contribution will dominate the ATI spectrum near the
cutoff, although the integrated yield of the 1st return in the
high recollision energy regime in Fig. \ref{FIG.3}(c) is
considerably larger than that of the 3rd return.

It is noteworthy that laser fields with various wavelengths from
infrared to mid-infrared are applied in the LIED scheme to extract
time-resolved dynamics of molecules
\cite{MmSCI2008,BlagaNAT2012,XuNC2014,PullenNC2016,MgNC2015,BwSCI2016}.
In order to give a more complete assessment of the Coulomb effect
in the LIED approach, in Fig. \ref{FIG.4} we present the CTMC
simulated results
for Ar at wavelength of 3100 nm and intensity of $1.25\times10^{14}%
$W/cm$^{2}$, which is another kind of typical laser pulse applied
in LIED \cite{BwSCI2016}. Overall, the distributions at 3100 nm in
Fig. \ref{FIG.4} are similar to the results at 800 nm. As shown in
Fig. \ref{FIG.4}(a), the 3rd-return-recollision trajectories
dominate the high energy part of the ATI spectrum. In Figs.
\ref{FIG.4}(c) and (d), similar to the case of 800 nm, the 2D
distribution of $s$ and $E_{r}$ for the 1st return distributes
uniformly in the impact parameter direction, while the
distribution for the 3rd return concentrates in the small impact
parameter regime. However, upon a careful inspection, there are
some differences between the 800 nm and 3100 nm cases. First, the
peak on the top of the 3rd return in Fig. \ref{FIG.3}(c)
disappears in Fig. \ref{FIG.4}(b). This causes the ATI spectrum
cutoff of the 3rd return obviously smaller than the 1st return. As
a consequence, a two-step, i.e., two-cutoff structure arises on
the spectrum in Fig. \ref{FIG.4}(a). It should be noted that this
two-cutoff structure may not be observed in experiment for two
reasons. (i) The signal of the second cutoff is too low for
experimental observation; (ii) The realistic cutoff in a quantum
system would not be as sharp as in the classical simulations,
which makes the two cutoffs hard to be distinguished. Second, the
difference between contributions from the 1st return and the 3rd
return to the spectrum at 3100 nm is much greater than that at 800
nm, which can be clearly seen by comparing Fig. \ref{FIG.4}(a)
with Fig. \ref{FIG.2}(c). These differences can be attributed to
the weaker effect of Coulomb field at 3100 nm. When wavelength
increases, the quiver distance increases quickly but the effective
radius of the Coulomb field keeps unchanged, thus the piece of the
trajectory affected by the Coulomb field decreases. In addition,
the electron momentum also increases and becomes harder to be
disturbed by the Coulomb field. Since the peak on the top of the
3rd return in Fig. \ref{FIG.3}(c) is induced by Coulomb field
effect, its absence at 3100 nm is not surprising. The difference
of relative contributions between the 1st and 3rd returns can be
interpreted with the help of the so-called defocusing effect
\cite{KBM2016,WangPRA2017}, namely, stronger focusing leads to
stronger deflection of the electron at the 1st return, which
causes the electron hard to come back to the core at the
subsequent returns. Since the Coulomb focusing effect is more
prominent for 800 nm, the contribution of the 3rd return is
suppressed due to the accompanying enhanced defocusing effect.

Based on the above, we believe that the 3rd-return-recollision
trajectories, instead of the 1st ones, should be used for analysis
in the LIED scheme. In this case, the imaging results will be
affected since the travel time for the 3rd-return-recollision
trajectories is about one optical cycle longer than that of the
1st return. Then the extracted result is actually the molecular
structure at the time about $7/4$ T instead of $3/4$ T after
ionization, which is crucial if LIED is applied to investigate the
time evolution of the molecular structure after ionization.

In conclusion, we perform a joint investigation on the dependence
of ATI spectrum on the pulse duration theoretically and
experimentally. The results indicate that the
3rd-return-recollision trajectories dominate the high energy part
of the ATI spectrum due to the ionic Coulomb field effect, which
invalidates the key assumption of SFA applied in LIED. The
incident (return) electron beams associated with different returns
show distinct characteristics in the presence of the Coulomb
field. Compared with the 1st return, the 3rd return generates an
electron beam with a much smaller diameter and a much higher
intensity, thus providing more high energy photoelectrons.
Moreover, it is found that the Coulomb field will also increase
the cutoff energy of the 3rd-return-recollision trajectories,
although this effect will become weaker with increasing
wavelength. The above Coulomb effect will change the results
extracted from the LIED approach and should be taken into account
in current imaging schemes based on recollision physics.

This work is supported by the National Key program for S\&T Research and
Development (No. 2019YFA0307700 and No. 2016YFA0401100), the National Natural Science Foundation
of China (Nos. 11504215, 11774387, 11874246, 11834015, 11974383), the Strategic Priority Research Program of the Chinese Academy of Sciences (No. XDB21010400), and the Science and Technology Department of Hubei Province (No. 2019CFA035).

\end{document}